# THE EFFECT CLAY ON THE MECHANICAL PROPERTIES OF EPOXY RESINS SUBJECTED TO HYGROTHERMAL AGEING

Salah U. Hamim[1] & Raman P. Singh[1]

[1]Mechanical and Aerospace Engineering

Oklahoma State University

Helmerich Research Center, 700 North Greenwood Ave, Tulsa, OK 74106, USA

## ABSTRACT

Epoxy polymers are an important class of material for use in various applications. Due to their hydrophilic nature, epoxy resins tend to absorb moisture. Absorption of moisture degrades the functional, structural and mechanical properties. For polymers, it is known that moisture absorption can lead to both reversible and irreversible changes. In this study, the combined effect of moisture and elevated temperature on the mechanical properties of Epon 862 and its nanocomposites were investigated. The extent of permanent damage on fracture toughness and flexural properties of epoxy, due to the aggressive degradation provided by hygrothermal ageing, was determined by drying the epoxy and their clay/epoxy nanocomposites after moisture absorption. From the investigation it was found that, hygrothermal ageing has detrimental effect on mechanical properties of both neat polymer and clay/polymer nanocomposite. Significant permanent damage was observed for fracture toughness and modulus, while the extent of permanent damage was less significant for flexural strength.

## 1. INTRODUCTION

Thermosetting polymers, such as epoxy resins and unsaturated polyesters are highly cross-linked and are very important class of advanced materials. Their main distinction from other types of polymers lies in their densely cross-linked molecular structure. This cross-linking leads to a number of favorable thermal and mechanical properties including high strength and modulus, high creep resistance, high glass transition temperature, low shrinkage, good resistance to chemicals, etc. These properties in conjunction with ease of processing have made epoxy resins an attractive choice for use in many engineering components and structures. They have found huge applications in aerospace, automotive, packaging, coating and microelectric industries. In recent years, researchers have developed and investigated polymer nanocomposites based on a wide variety of nano-scale fillers including clay particles (layered silicates) [1], $TiO_2$ particles, carbon nanotubes, etc.

Effect of clay loading on mechanical properties of epoxy polymers has been studied previously to a certain extent. Zhao and Li [2] reported tensile strength and modulus decreased for both neat epoxy and Nanocomposites upon moisture absorption, while the tensile strain increased significantly for moisture absorbed samples. Similar observation of strength and modulus decreasing upon moisture absorption has been reported by Glaskova and Aniskevich [3]. Wang et al. [4] investigated the effect of hydrothermal effects on mechanical properties such as tensile strength, modulus and fracture toughness with immersion duration. For DGEBA epoxy systems, fracture toughness and modulus was not influenced much with immersion time, while strength decreases for nanocomposites. According to the study conducted by Buck et al. [5] at elevated temperature, combination of moisture and sustained load can significantly reduce ultimate tensile strength of E-glass/vinyl-ester composite materials. A study on elastic modulus of epoxy polymer after a absorption-desorption cycle showed recovery of property from wet condition, although modulus remains at a lower value than as-prepared samples for lower filler volume. For higher volume fraction of reinforcement, elastic modulus improves to a value which is more than the elastic modulus of as-prepared samples [6]. Ferguson and Qu [7] also reported recovery of elastic properties from moisture saturated state after desorption cycle. However, DeNeve and Shanahan [8] did not observe any recovery of elastic modulus after an absorption-desorption cycle at elevated temperature. Most of the research on polymer-clay nanocomposites has focused on investigating the effect of various parameters on mechanical properties such as modulus and strength. Although fracture toughness is a very important property for these nanocomposites to be used in various structural applications, it has not been studied adequately and the results reported in the literature are inconclusive. Furthermore, the effects of moisture absorption on fracture toughness of polymer-clay Nanocomposites has not been studied extensively and to our best knowledge no study was conducted to investigate the recovery of fracture and flexural properties after a absorption-desorption cycle. Durability of polymer/clay nanocomposites are still needed to be studied in depth, particularly for hygrothermal aging in which the degradation of the mechanical properties and loss of integrity of these nanocomposites occur from the simultaneous action of moisture and temperature.

This study on epoxy/clay nanocomposite is designed to investigate the effect of hygrothermal ageing on mechanical properties of these nanocomposites. A drying cycle is employed to quantify the recovery of the properties after hygrothermal aging. This would be helpful to understand the extent of permanent damage occurred by the combined action of elevated temperature and moisture. Fracture toughness, flexural strength and modulus are the properties that were studied. Two structurally different clays are used as reinforcement in epoxy matrix in order to aid comparison.

## 2. EXPERIMENTATION

### 2.1 Materials

The epoxy resin used for this study is diglycidyl ether of bisphenol F (Epon 862, Miller-Stephenson Chemical Company, Inc., Dunbury, Connecticut, USA). The curing agent used for this resin system is a moderately reactive, low viscosity aliphatic amine curing agent (Epikure 3274, Miller-Stephenson Chemical Company, Inc., Dunbury, Connecticut, USA). For mixing the clay with epoxy, a high-speed shear disperser was used (T-25 ULTRA TURRAX with SV 25 KV25 F dispersing element, IKA Works Inc., Wilmington, North Carolina, USA). The clay used for this study is synthetic fluorinated mica modified with di-methyl di-tallow quaternary ammonium (Somasif MAE, Co-op Chemicals, Japan).

## 2.2 Sample preparation

Epoxy was preheated to 65°C before desired amount clay was introduced and mixed using mechanical mixer for 12 hours. To reduce the viscosity of the mixture and facilitate mixing, temperature was maintained at 65°C for the entire duration of mixing using a hot plate. To remove entrapped bubble from the mixture degassing was done for around 30 minutes. Bubble free mixture of clay and epoxy was then shear mixed at 13000 rpm for 30 minutes. During this process, temperature was maintained at 65°C using an ice bath. Subsequently, the mixture was then degassed until it was completely bubble-free. Curing agent was added to the mixture at 100:40 ratio and carefully hand mixed to avoid introduction of any air bubble. After it was properly mixed, the final slurry containing epoxy and clay was poured in to an aluminum mold and cured at room temperature for 24 hours followed by post-curing at 121°C for 6 hours. The final outcome has a nominal dimension of 7 in. x 6 in. x 0.25/0.125 in. The weight fraction of the clay was varied from 0.5 - 2.0 % to study the influence of clay on mechanical properties' of nanocomposites after degradation.

## 2.3 Environmental preconditioning

After specimens were cut into final required dimension, they were subjected to degradation. Specimens from each nanocomposite was taken and submerged in boiling water for 24 hours. It was experimentally found that 24 hours of time was enough to ensure saturation of water uptake into the specimens. Water saturated specimens were dried in an oven at 110°C for 6 hours to remove void-filling moisture from the samples leaving only permanent degradation in form of bonded water.

## 2.4 Fracture toughness determination

Mode-I fracture toughness was determined by Single Edge Notch Bend (SENB) test as per the ASTM D-5045 on Universal Testing Machine (Instron 5567, Norwood, MA) in a displacement-controlled mode with fixed crosshead speed of 10 mm/min. Nominal dimension for the SENB test samples were 2.65 inch x 0.6 inch x 0.25 inch. For fracture toughness determination a notch was created using precision diamond saw (MK-370, MK Diamond Products Inc., Torrance, California, USA). A sharp pre-crack with ratio of 0.45<a/W<0.55 was created by tapping a fresh razor blade into the notch. At least 5 specimens were tested for every condition and nanocomposite. Fracture toughness for the specimens was calculated in terms of critical stress intensity factor following the equations provided by the ASTM standard. The crack length was measured using an Optical Microscope (Nikon L150) which has a traveling plate with graduations.

## 2.5 Flexural strength and modulus determination

Flexural Modulus and Strength was determined using Three Point Bend (3PB) test according to ASTM D790 on Universal Testing Machine (Instron 5567, Norwood, MA). The nominal dimension for the flexural test specimens were 2.2 in x 0.5 in x 0.125 in. The crosshead speed for the test was calculated using equations from the ASTM standard. The crosshead speed was found to be 1.35 mm/min. Flexural strength was calculated using the maximum load on the load-deflection curve and flexural modulus was calculated using the slope of the tangent to the initial straight line portion of the load-deflection curve.

# 3. RESULTS

## 3.1 Fracture toughness

The critical stress intensity factor as a function of clay loading for Somasif clay/epoxy nanocomposites are shown in figure 1. In this figure, fracture toughness of nanocomposites subjected to different conditioning was plotted and $K_{Ic}$ values for as-prepared samples were also listed as a reference. It is found that, incorporation of clay into the epoxy matrix leads to increased toughness for nanocomposites compared to the neat polymer. For 0.5wt% addition of clay, critical stress intensity factor, $K_{Ic}$ increased 14% compared to the neat polymer. In case of higher clay loading the fracture toughness does not necessarily improve compared to 0.5wt% clay/epoxy nanocomposite. As higher clay content can lead to improper exfoliation and agglomeration, it is possible to assume that, the positive reinforcing effect of clay particles are neutralized by the agglomerates in these nanocomposites.

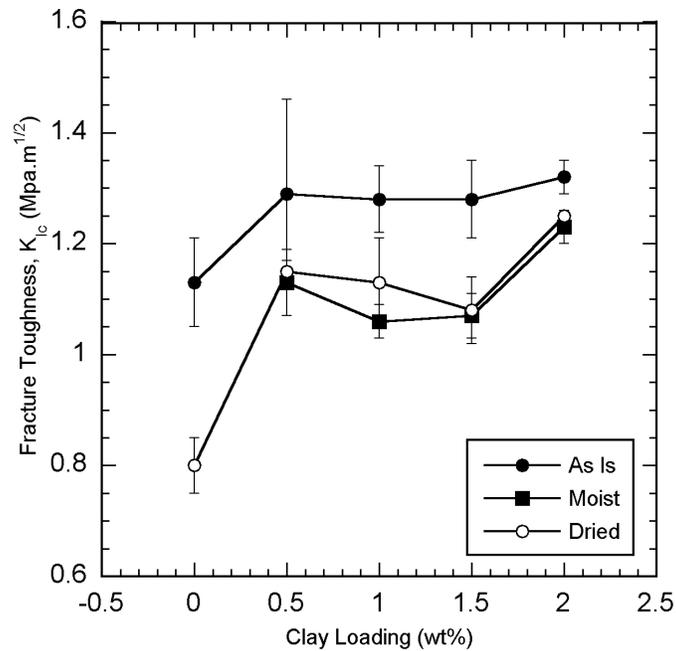

**Figure 1: Fracture toughness as a function clay loading percentage**

For neat polymer, the moisture absorbed specimens showed excessive plasticization. As linear elastic fracture mechanics does not hold for viscoelastic materials, it was not possible to determine the fracture toughness of moisture absorbed neat polymer using Single-edge-notch-bend test. In case of clay/epoxy nanocomposites, moisture absorbed specimens showed reduction in critical stress intensity factor compared to the as-prepared samples. For 0.5wt% clay/epoxy nanocomposite, the fracture toughness reduced 12.4% from the as-prepared nanocomposite. Similar reduction of fracture property was observed for other clay percentages. Once absorbed in polymer system water acts as plasticizer; however, no plasticization effect was observed for clay/epoxy nanocomposites after moisture absorption.

Dried neat polymer, free of void-filling or free water showed serious degradation in terms of fracture toughness. 29.2% reduction in fracture toughness was observed for dried neat polymer compared to the as-prepared neat polymer. Dried clay/epoxy nanocomposites did not show any significant recovery of fracture toughness from the moisture absorbed state. However, it can be noticed that the extent of permanent damage to the fracture property was much less in nanocomposites than in neat polymer. For 0.5wt% clay, same conditioning reduced the fracture toughness by only 10.85% from the original value.

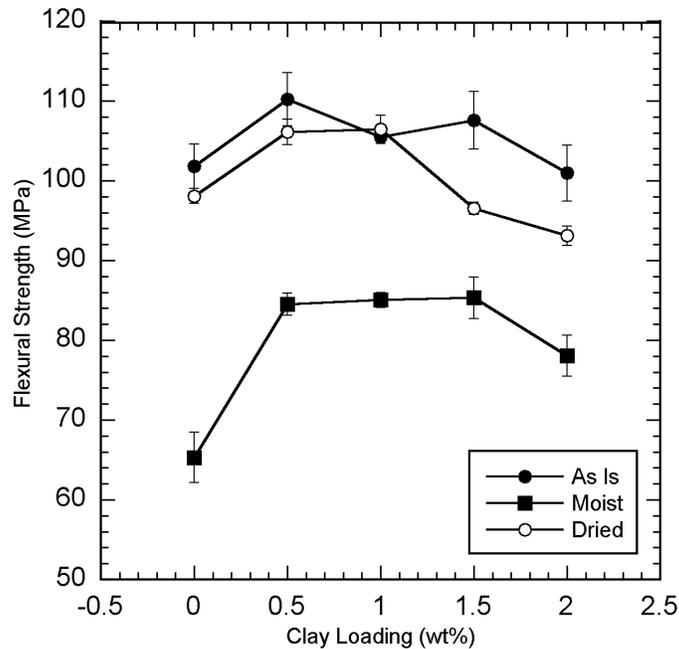

**Figure 2: Flexural strength as a function clay loading percentage**

## 3.2 Flexural strength and modulus

Flexural strength and modulus were determined for the as-prepared and conditioned specimens as a part of the study. Figs. 2 and 3 show the flexural strength and modulus as a function of clay loading, respectively. From the figures it can be seen that, absorption of moisture leads to severe degradation of the flexural properties. For neat polymer, reduction of flexural strength and modulus were 35.86% and 20%, respectively. These properties were less severely affected for clay/epoxy nanocomposites compared to the neat polymer. A 23.3% and 11.1% reduction of flexural strength and modulus were observed for 0.5wt% clay/epoxy nanocomposites, respectively.

After drying, both the properties showed recovery from the moisture absorbed state. Flexural strength recovered almost fully and the permanent damage was in the range of only about 5%. Flexural modulus also improved considerably after drying for neat polymer and lower clay loading nanocomposites. It is interesting to note that for higher clay loading the recovery in flexural modulus is not comparable with the recovery of lower clay loading nanocomposites' recovery of modulus.

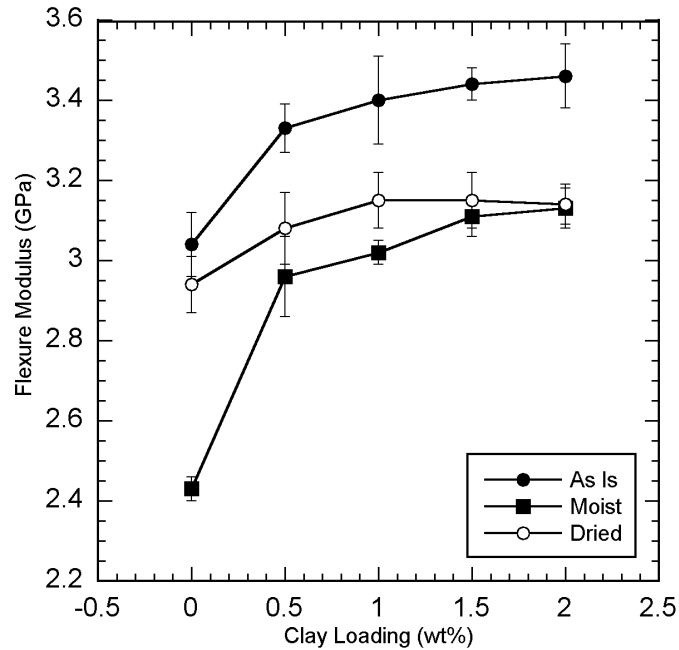

**Figure 3: Flexural modulus as a function clay loading percentage**

## 4. CONCLUSIONS

Fracture toughness, flexural strength and modulus were determined for two different clay/epoxy nanocomposites following the ASTM standards. The effect of hygrothermal ageing on the properties was investigated. After removing the free water by drying, the irreversible effect or the permanent damage due hygrothermal aging on the clay/epoxy nanocomposite systems were also studied. From the study it was observed that, all three properties were degraded due to hygrothermal ageing. However, incorporation of clay in epoxy matrix has positive effects to some extent. For example, permanently damaged nanocomposites provided more toughening than pristine polymer. It was also observed from the collected data that, properties were less severely degraded for clay/epoxy nanocomposites compared to neat epoxy samples.